\documentclass[a4paper, times, 10pt,twocolumn]{article}
\usepackage[top=4.9cm,bottom=3.7cm,left=1.5cm,right=1.5cm]{geometry}
\usepackage{ICMLC}
\usepackage{times}
\usepackage{graphicx}
\usepackage{indentfirst}
\usepackage{latexsym}
\usepackage{enumitem}
\usepackage{color}
\usepackage{bm}
\usepackage{version}
\excludeversion{hide}
\usepackage{amssymb,mathtools,times}
\newcommand{\blambda}{{\bm{\lambda}}}
\newcommand{\e}{{\mathrm e}}
\newcommand{\im}{{\mathrm i}}
\newcommand{\dd}[1]{\, {\mathrm d}{#1}}
\newcommand{\bdm}{\begin{displaymath}}
\newcommand{\edm}{\end{displaymath}}
\newcommand{\cond}{\,|\,}
\newcommand{\define}{\, := \,}

\newcommand{\btheta}{{\bm{\theta}}}

\newcommand{\Rf}{{\mathbb{R}}}

\usepackage[margin=8pt,font=footnotesize,labelfont=bf]{subcaption}

\pdfpagewidth=\paperwidth
\pdfpageheight=\paperheight
\begin{document}

\title{THE SCATTERING TRANSFORM NETWORK \\ WITH GENERALIZED MORSE WAVELETS AND \\
  ITS APPLICATION TO MUSIC GENRE CLASSIFICATION}

\author{\bf{\normalsize{WAI HO CHAK${^1}$, NAOKI SAITO${^1}$, DAVID WEBER${^1}$}}\\ 
\\
\normalsize{$^1$ Graduate Group in Applied Mathematics, University of California, Davis, USA. }\\
\normalsize{E-MAIL: wchak@ucdavis.edu, saito@math.ucdavis.edu, dsweber@math.ucdavis.edu  }\\
\\}

\maketitle \thispagestyle{empty}

\begin{abstract}
  {We propose to use the Generalized Morse Wavelets (GMWs) instead of
    commonly-used Morlet (or Gabor) wavelets in the Scattering Transform
    Network (STN), which we call the GMW-STN, for signal classification
    problems. The GMWs form a parameterized family of truly analytic wavelets
    while the Morlet wavelets are only approximately analytic. The analyticity
    of underlying wavelet filters in the STN is particularly important for
    nonstationary oscillatory signals such as music signals because it improves
    interpretability of the STN representations by providing multiscale
    amplitude and phase (and consequently frequency) information of input
    signals.
    We demonstrate the superiority of the GMW-STN over the conventional STN
    in music genre classification using the so-called GTZAN database. 
    Moreover, we show the performance improvement of the GMW-STN by increasing
    its number of layers to three over the typical two-layer STN.}
\end{abstract}

\begin{keywords}
  {Generalized Morse Wavelets; Analytic Wavelet Transform; Scattering Transform;
    Music Genre Classification}
\end{keywords}

\Section{Introduction}
\label{sec:intro}

A Convolutional Neural Network (CNN), in particular, its ``deeper'' version,
a Deep Neural Network (DNN), has been shown to be effective in the extraction of
hierarchical features for many applications with large training
datasets~\cite{lecun2015deep}. However, there is a lack of interpretability of
the DNN outputs. Also, the DNN does not demonstrate a good performance without
having a large dataset due to model overfitting. 

On the other hand, the \emph{Scattering Transform Network} (STN), which has a
similar architecture with the CNN, can generate more interpretable
representations of input data. In addition, the STN works reasonably well on
smaller datasets because the convolutional filters are pre-constructed using
the established techniques from signal processing. Originally, Mallat proposed
the STN to connect the wavelet theory with the CNN, and showed that the STN
generates a quasi-translation-invariant signal representation with a cascade of
wavelet filtering and modulus nonlinearities~\cite{mallat2012group}.
B\"olcskei and Wiatowski showed that with increasing depth,
the STN achieves better translation invariance~\cite{WIATOWSKI-BOLCSKEI-DEEP-THEORY-IT}.
Numerically, Bruna and Mallat illustrated the capability of the STN for texture
image classification ~\cite{bruna2011classification, bruna2013invariant}.

Most of these previous works on the STN used the Morlet (or Gabor) wavelet
filters~\cite[Sec.~4.3]{MALLAT-BOOK3}, which are only \emph{approximately}
analytic. The analyticity of the wavelet filters is quite important
for input signals that are nonstationary and oscillatory because it allows
us to represent them in terms of amplitude, phase, and frequency
in a multiscale manner~\cite{LILLY-OLHEDE-1}.

In this paper, we propose to use the so-called \emph{Generalized Morse Wavelet}
(GMW) filters~\cite{olhede2002generalized, LILLY-OLHEDE-3} as the wavelet
filters in the STN instead of the commonly-used Morlet wavelet filters.
The GMW provides a parameterized family of \emph{truly} analytic wavelets,
and adopting this in the STN framework should give us better performance
in classifying input signals and interpreting their STN representations.

A major application of our proposed method is music genre classification.
The input data here are recorded digital music signals,
which are nonstationary and quite oscillatory. Hence, we should be able to
see clear advantages of using the GMWs over the Morlet wavelets.

\begin{hide}
Our contributions in this paper are:
\begin{enumerate}[topsep=0pt,itemsep=-1ex,partopsep=1ex,parsep=1ex]
\item incorporating the GMW in the STN for the analysis of 1D nonstationary
  signals including the music signals;
\item demonstrating superior performance of the GMW in the STN over
  the Morlet wavelet for classifying music signals into genres.
  The result can be mathematically explained by the analyticity of the wavelet.
\item providing the better classification performance of the music genres
  when the STN is deeper. We provide the interpretation of the output STN coefficients from the music signals whereas the conventional deep learning approaches cannot.
\end{enumerate}
\end{hide}

\section{Wavelet and Scattering Transforms}
\label{sec:WSTs}
In this section, we will introduce the mathematical methods that are essential
to help us process the music/audio signals. These also serve as the foundation
to understand the motivation of our new method for music signal classification.

\subsection{Generalized Morse Wavelets (GMWs)}
\label{sec:GMWs}
Let us review the concept of \emph{analytic wavelets} for signal analysis.
We will also describe the properties of the \emph{Generalized Morse Wavelets}
(GMWs) and the \emph{Continuous Wavelet Transform} (CWT), both of which are
crucial for the STN in Section~\ref{sec:ST}.

A \emph{(mother) wavelet} $\psi(t) \in L^2(\Rf)$ is a function whose dilated and
translated versions provide a method to perform localized time-frequency
analysis of nonstationary oscillatory signals,
such as audio and music signals~\cite[Sec.~4.3]{MALLAT-BOOK3}.
A wavelet is said to be \emph{analytic} if it is complex-valued with vanishing
support on negative frequencies; see, e.g., \cite[Sec.~4.3]{MALLAT-BOOK3}.
The \emph{Continuous Wavelet Transform} (CWT) of a signal $g(t) \in L^2(\Rf)$
with respect to the mother wavelet $\psi$ is given by 
\begin{equation}
W_\psi g(a, b) \define \dfrac{1}{\sqrt{a}} \int_\Rf g(t) \overline{\psi\left(\dfrac{t - b}{a} \right)} \dd{t} ,
\end{equation}
for any $a \in \Rf_+ \define \{ t \in \Rf \cond t > 0 \}$, $b \in \Rf$.
The CWT is called the \emph{Analytic Wavelet Transform} (AWT)
when $\psi$ is analytic.

A follow-up question that we have to address is which analytic wavelet is
suitable in practice. The Morlet wavelet was extensively used in the STN
literature and software implementation. 
However, Lilly and Olhede~\cite{LILLY-OLHEDE-3, LILLY-OLHEDE-2}
demonstrated numerically that even small leakage to negative frequencies in
the Morlet wavelet can lead to abnormal transform phase variation.

On the contrary, the \emph{Generalized Morse Wavelets} (GMWs) is a promising
superfamily of \emph{truly analytic} wavelets~\cite{olhede2002generalized, LILLY-OLHEDE-3}.
In the frequency domain, the GMW is defined as
\begin{equation}
\Psi_{\beta, \gamma}(\omega) \define \int_\Rf \psi_{\beta, \gamma}(t) \e^{-\im\omega t} \dd{t} = H(\omega) \alpha_{\beta, \gamma} \omega^\beta \e^{-\omega^\gamma} ,
\end{equation}
where $\beta>0, \gamma>1$ are two main parameters, $\alpha_{\beta, \gamma}$
is a normalization constant, and $H(\omega)$ is the Heaviside step function.
The parameters $\beta$ and $\gamma$ control the time-domain and frequency-domain
decay, respectively.
The peak frequency $\omega_{\beta, \gamma} \define (\beta / \gamma)^{1 / \gamma}$ is
the frequency at which the derivative of $\Psi_{\beta, \gamma}$
vanishes~\cite{LILLY-OLHEDE-3, LILLY-OLHEDE-2}.
The numerical implementation and experiment by
Lilly and Olhede~\cite{LILLY-OLHEDE-3, LILLY-OLHEDE-2} illustrate that the GMWs
are supported only on positive frequencies unlike the Morlet wavelets.
Thus the statistical properties will not be destroyed due to departures from
analyticity if one adopts the GMWs.

\subsection{Scattering Transform Network (STN)}
\label{sec:ST}
\begin{figure}
\centering
\includegraphics[scale = 0.39]{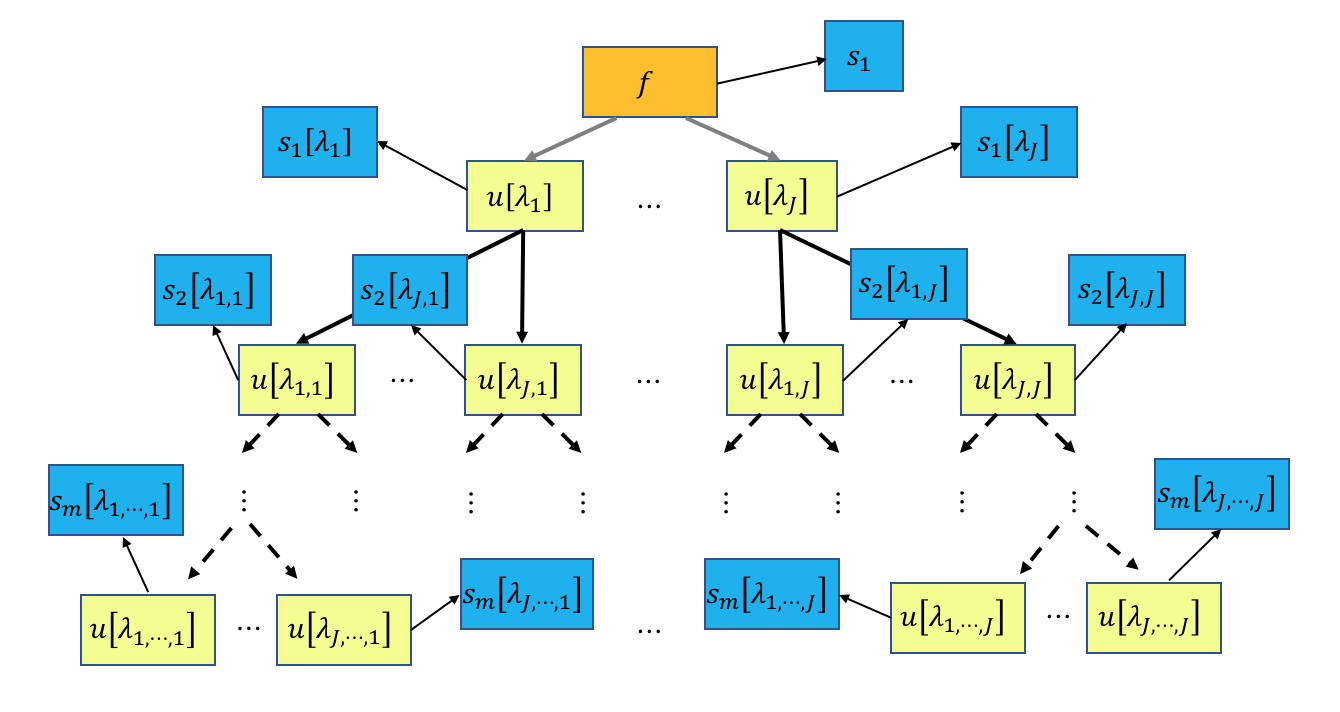}
\caption{A typical STN architecture}
\label{fig:st}
\end{figure}

The architecture of the \emph{Scattering Transform Network} (STN) is a
tree-like analog of a convolutional neural network. Figure~\ref{fig:st}
illustrates its typical architecture.

At the $m$th layer of the STN, we denote
$\lambda_m = (j_m-J_m)/Q_m$, $j_m \in \{0, 1, \ldots, J_m\}$, the index for a
multiscale wavelet filter where $Q_m > 0$ is the so-called \emph{quality factor}
and $2^{J_m/Q_m}$ is the largest scale of interest at the $m$th layer.
Hence, $j_m=0$ corresponds to the coarsest scale/lowest frequency band
whereas $j_m=J_m$ corresponds to the finest scale/highest frequency band.
By dilating a mother wavelet $\psi$, we can generate multiscale wavelet filters,
i.e., 
\begin{equation}
  \psi_{\lambda_m}(t) \define 2^{\lambda_m} \psi\left(2^{\lambda_m} t\right)
  \Leftrightarrow \Psi_{\lambda_m}(\omega) = \Psi\left(2^{-\lambda_m}\omega\right) ,
\end{equation}
where we assume $\psi_{\lambda_m} \in L^1(\Rf) \cap L^2(\Rf)$.
These 
correspond to receptive fields of a CNN~\cite{WIATOWSKI-BOLCSKEI-DEEP-THEORY-IT}.

Let $f \in L^2(\Rf)$ be an input signal of interest.
We define a \emph{contraction} operator $M_m$ which is
\emph{Lipschitz continuous},
and satisfies the condition $M_m f(t) = 0 \Rightarrow f(t) = 0$.
One popular choice of $M_m$ is the \emph{modulus} operator, i.e.,
$M_m f(t) \define | f(t) |$, which we will use in our experiments in
Section~\ref{sec:music}.
Let $\Lambda_m$ be the set of indices $\{\lambda_m\}$ for the $m$th layer.
Each internal layer operator $U_m: \Lambda_m \times L^2(\Rf) \rightarrow L^2(\Rf)$ is defined by three operations: 1) the (linear) wavelet transform; 2) the
contraction operator; 3) subsampling:
\begin{equation}
U_m[\lambda_m] f(t) \define M_m \left(f \ast \psi_{\lambda_m}\right)(r_m t) ,
\label{eqn:u1-def}
\end{equation}
where $r_m \geq 1$ is the \emph{subsampling rate}.
Therefore, there is a \emph{path} of indices
$\blambda \in \Lambda_m \times \cdots \times \Lambda_1$ such that
\begin{equation}
U[\blambda] f(t) \define U_m[\lambda_m] U_{m-1}[\lambda_{m-1}] \cdots U_1[\lambda_1] f(t) .
\label{eqn:u2-def}
\end{equation}

We define the operator $S_m$ for each layer $m$ to generate the robust
multiscale features of the input signal $f(t)$, which we also call
``coefficients'' or ``representations.''
\begin{eqnarray}
  \label{eqn:sm-def}
  S_m[\blambda]f(t) \define \left(\varphi_m \ast U[\blambda] f\right) \left(r'_m t \right),
\end{eqnarray}
where $\varphi_m$ represents the averaging function, or father wavelet at a
certain scale corresponding to the mother wavelet $\psi_m$. After the averaging
stage, we can subsample again at rate $r'_m \geq 1$ since the averaging wavelet
is a lowpass filter. Note that in particular, for layer $m=0$,
we have $S_0[\emptyset]f(t) \define (\varphi_0 \ast f)(r'_0 t)$.

Note that the GMWs were originally implemented in MATLAB\textregistered\
and disseminated as the \textsf{JLAB} package~\cite{LILLY-OLHEDE-2}.
We implemented the STN with the option of using either the Morlet wavelets or
the GMWs~\cite{ScatteringTransform} in the Julia programming language~\cite{JULIA},
and we refer to the STN with the latter option as the \emph{GMW-STN}.
The doublet of parameters $(\beta, \gamma)$ of the GMWs in our implementation
is set as $(4, 2)$ considering the balance of the time-frequency decays.


\section{Music Genre Classification using the GMW-STN}
\label{sec:music}
Categorizing recorded music signals into different genres such as classical,
country, hiphop, jazz, pop and so on is a difficult task, in part because
classification of music genres by human judgment can be subjective and
ineffective. In such a classification problem, several CNN-based methods
have been proposed, e.g., 
\cite{allamy20211d} among others. Although these methods offered some escape
routes, they cannot completely escape from the following two fundamental
problems (as discussed in Introduction) when they are applied to music signal
databases of relatively small size:
1) model overfitting; and 2) interpretability of the intermediate signal
representations within CNNs.
\begin{hide}
The 2D convolutional recurrent neural network (CRNN) was proposed by Choi et al.~\cite{choi2017convolutional} for categorizing music.
The local deep feature extracted by the convolutional neural network (CNN) and the summarization of the extracted features by the recurrent neural network are shown to achieve effective music classification. Allamy and Koerich ~\cite{allamy20211d} also mentioned the effectiveness of 2D CNN ~\cite{costa2017evaluation} and suggested the use of 1D residual CNN in the GTZAN dataset~\cite{tzanetakis2002musical} to automatically classify music genres.  Data augmentation may be able to generate enough data and improve the performance ~\cite{allamy20211d}, but the method heavily depends on the quality of the original dataset and whether the augmentation technique is appropriate. Still, this does not help us understand how to interpret the result from the deep network. It is of interest to know how the features are interpreted so that we can explain the high classification rate with a better understanding of the characteristics of different genres.
\end{hide}
Hence, such music genre classification with a music signal database of small
size is an ideal application to demonstrate the advantage of the STN, in
particular, the GMW-STN.


\subsection{The GTZAN database and data preparation}
\label{sec:data}
In our experiments, we used the so-called GTZAN
database~\cite{tzanetakis2002musical} contains 1,000 audio/music tracks
each of which is 30 second long and was sampled at 22,050Hz.
The tracks are evenly distributed into ten music genres:
blues; classical; country; disco; hiphop; jazz; metal; pop; reggae; and rock.
For each music genre, the 100 tracks were recorded under different conditions. 

In our experiments, we split each 30-second track into a set of 15 overlapping
segments each of which is 5 second long. Let $k$ denote the index of such a
music segment. The time interval (indexed by samples) of the $k$th music segment
is $[ kL/3 + 1, kL/3 + L]$ for $k=0:14$, where $L=22050 \cdot 5$ samples;
that is, the hop size is $L/3$, i.e., the two adjacent segments have $2/3 \cdot
5 \approx 3.33$ second overlap.

In our three-layer STN, we set the subsampling rates $r_m = 8$ and $r'_m = 32$,
which generated the following size of the STN coefficients for each input
signal of length $110,250$: the 0th layer: $3,445$; the 1st layer: $(431, 33)$;
the 2nd layer: $(54, 14, 33)$; and the 3rd layer: $(7, 10, 14, 33)$.
We also set the quality factors as $(Q_1, Q_2, Q_3) = (8, 4, 4)$, and
$(J_1, J_2, J_3)=(32, 13, 9)$.
These numbers mean that we used $33$ scales of the form $2^{(32-j_1)/8}$ with
$j_1=0:32$ in the 1st layer, $14$ scales of the form $2^{(13-j_2)/4}$ with
$j_2=0:13$ in the 2nd layer, and $10$ scales of the form $2^{(9-j_3)/4}$ with
$j_3=0:9$ in the 3rd layer.
The first numbers, $3,445$, $431$, $54$, and $7$, are the size of the
output coefficients in each path in the respective layers.

\begin{hide}
\begin{table}[h]
\caption{Output dimensions of STN for each music track.}
 \begin{center}
 \begin{tabular}{|l|l|}
  \hline
  {\bf Layer index} & {\bf Length} \\
  \hline\hline
  Input  & 110,250 \\
  \hline
  1  & (431, 33) \\
  \hline
  2  & (54, 14, 33) \\
  \hline
  3  & (7, 10, 14, 33) \\
  \hline
 \end{tabular}
\end{center}
 \label{tab:out_dim}
\end{table}
\end{hide}

\subsection{Interpreting music features in different layers}
\begin{figure*}
\begin{subfigure}{0.33\textwidth}
  \centering\includegraphics[scale = 0.285]{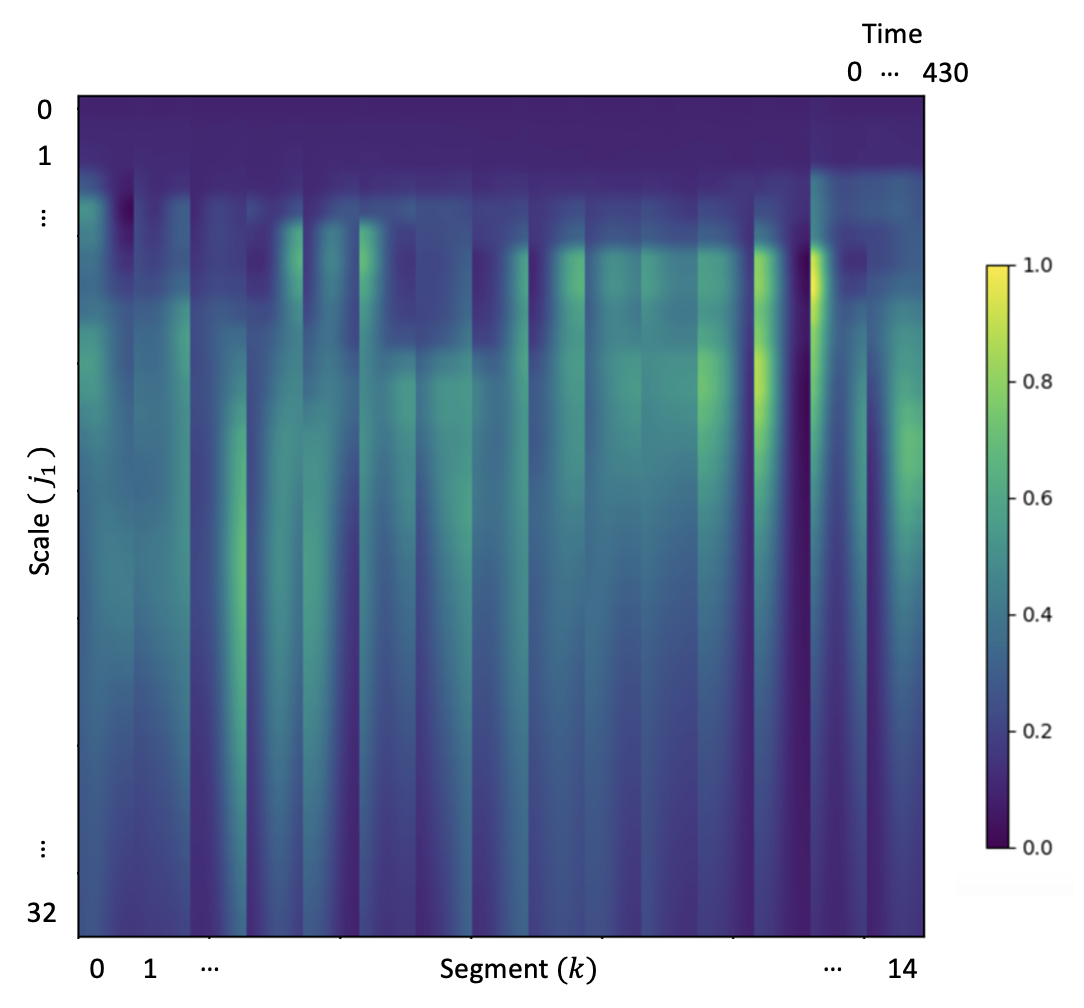}
  \caption{1st layer GMW-STN}
  \label{fig:1st}
\end{subfigure}
\begin{subfigure}{0.33\textwidth}
  \centering\includegraphics[scale = 0.285]{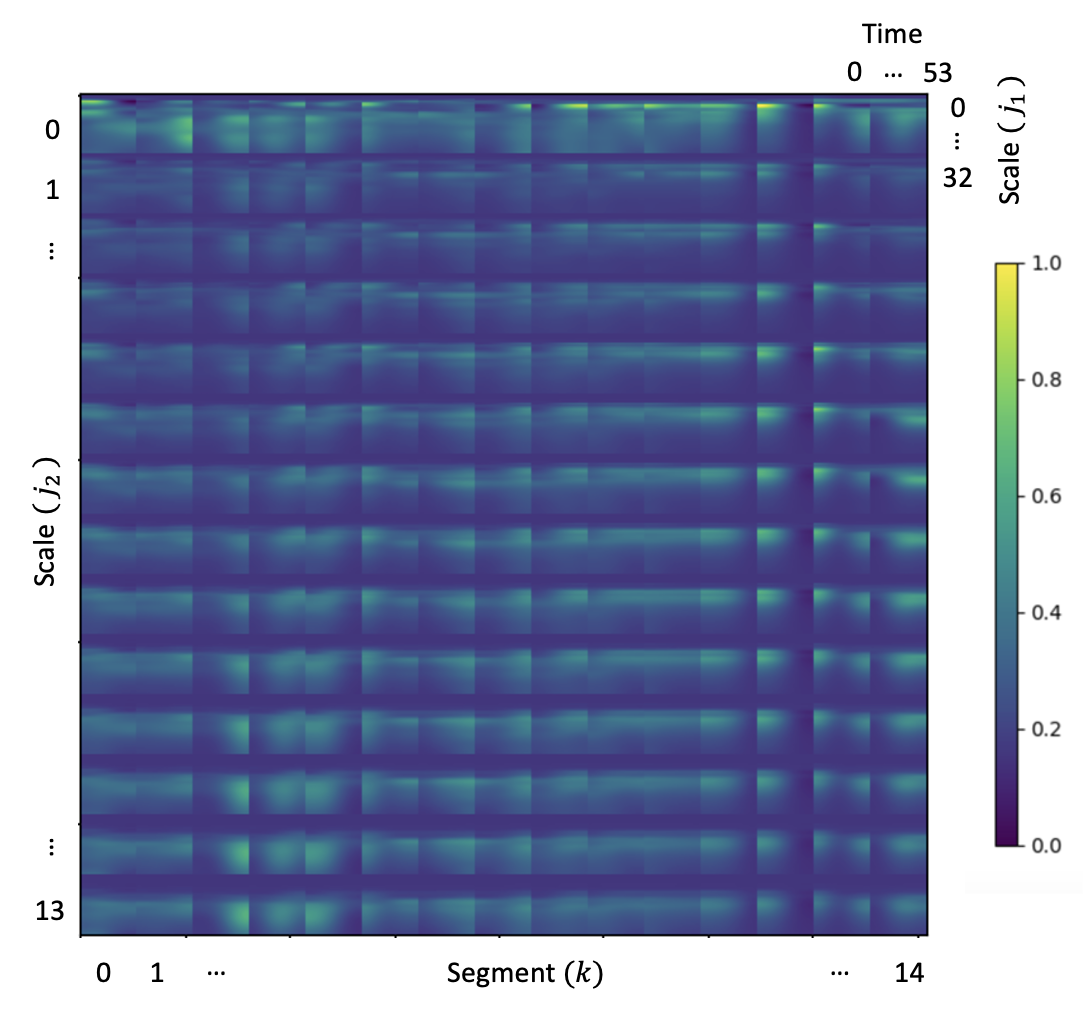}
  \caption{2nd layer GMW-STN}
  \label{fig:2nd}
\end{subfigure}
\begin{subfigure}{0.33\textwidth}
  \centering\includegraphics[scale = 0.285]{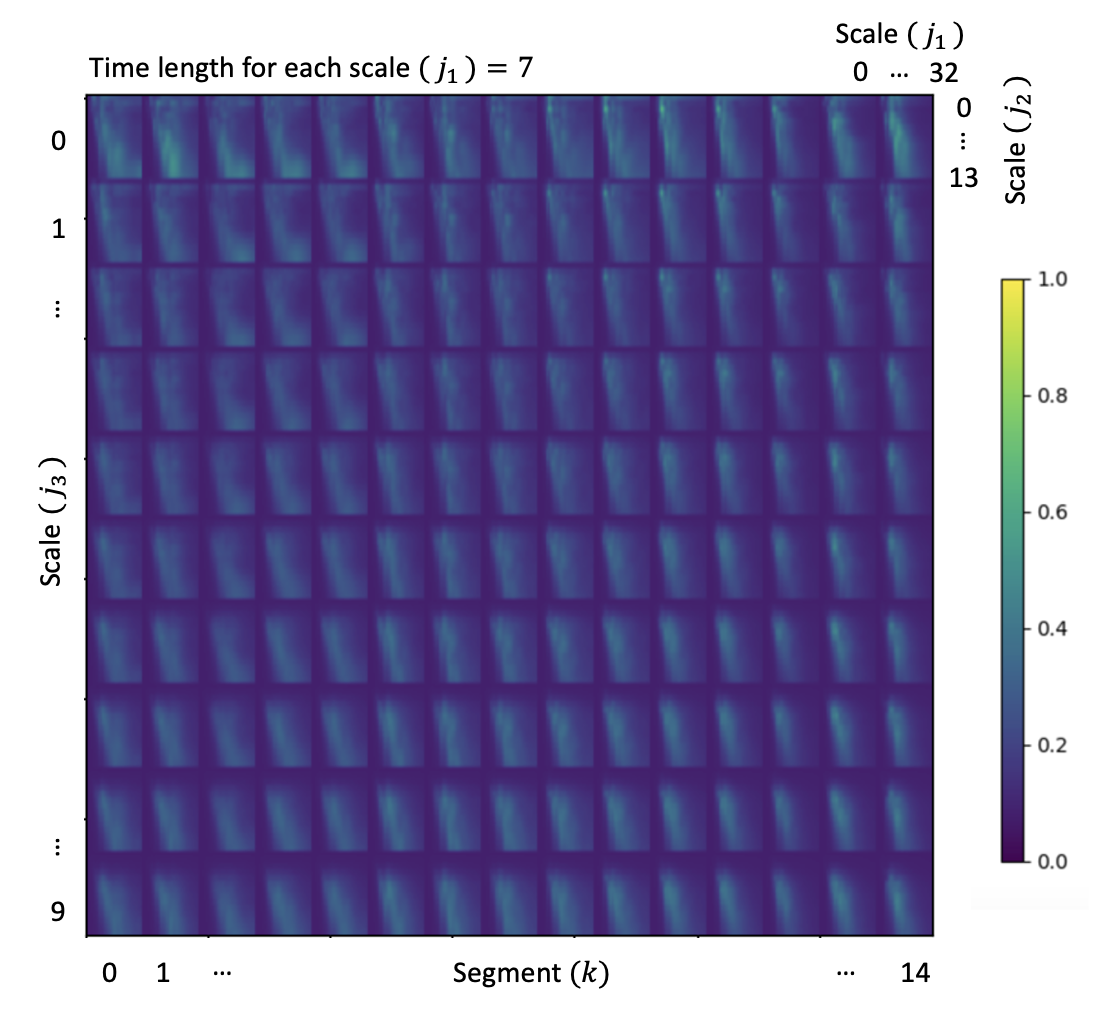}
  \caption{3rd layer GMW-STN}
  \label{fig:3rd}
\end{subfigure}
\caption{The normalized GMW-STN 1st, 2nd, and 3rd layer outputs from a jazz track}
\label{fig:1_layer}
\end{figure*}

In the 1st layer $m = 1$, we obtained a spectrogram-like STN output as shown in
Figure~\ref{fig:1st} for a jazz track. 
Most of the signal energy resides in low to medium frequency bands.  

The 2nd layer paths have two different frequency measurements, indexed by
$(j_2, j_1)$ where $j_2= 0:13$ and $j_1=0:32$.
Figure~\ref{fig:2nd} shows the 2nd layer outputs.
Each block in this subfigure represents the 2nd layer outputs corresponding
to the 2nd layer scale $j_2$ and the $k$th segment of this jazz track,
and the row index within each block corresponds to $j_1=0:32$.
The features corresponding to the 2nd layer scale $j_2 = 0$ has more
variations than those with the larger value of $j_2$ (i.e., finer scales).
The larger the value of $j_2$ is, the more stable and less noisy the outputs
become. 
The hierarchical features in the deeper layer of the STN become more invariant
(or stable) to the local deformations compared to the 1st layer features.
Yet there are many variations across different overlapping segments
in the 2nd layer output due to nonstationarity of the jazz music.

The 3rd layer paths add a third scale variation, indexed by $(j_3, j_2, j_1)$,
$j_3=0:9$.
Figure~\ref{fig:3rd} shows 3rd layer output. 
Each block in this subfigure represents the 3rd layer outputs indexed by the 3rd
layer scale $j_3$ and the $k$th segment of this jazz track,
and the row index within each block corresponds to $j_2=0:13$
while the columns within each block are organized by $j_1=0:32$ and for each
$j_1$, there are $7$ coefficients, i.e., the number of columns of each block
is $33 \times 7 = 231$.
The main advantage of using the 3rd layer is the increased quasi-translation
invariance. As in the 2nd layer, we observe more stable patterns when $j_3$
is larger. The major difference from the 2nd layer is that the patterns across
different overlapping music segments are more similar in the 3rd layer,
which seem to capture intrinsic features of music genre.
Thus, the hierarchical features are expected to have more discriminant power
to categorize music signals into different music genres. 



\subsection{Classification experiment settings}
We employed the three-fold cross validation scheme and repeated this 
ten times. In each experiment, the 1,000 music tracks were shuffled and grouped
into three folds of 340, 330, and 330 tracks. We used two folds for training and
one fold for testing. Then we iterated the process three times by permuting
the folds in each of those ten experiments. In other words, we ran the
classification experiments 30 times in total.
Each fold contains all music genres that are evenly distributed.
For instance, there are 34 music files for each music genre in the first fold
containing 340 files.
In each fold, we further split each music track into 15 segments as described
in Section~\ref{sec:data}. 

In the training stage, we extracted the STN outputs from all segments in the
training set. Then we compressed these outputs using the PCA implemented in
the \textsf{MultivariateStats.jl}. We chose the top $1,000$ principal components
based on the experimental performance. Then we fed these into a classifier.
In the testing stage, we assigned each input file a label based on the
majority vote among the labels of its 15 segments predicted by the trained
classifier.

In our numerical experiments, we mainly used the \emph{Support Vector Machine}
(SVM)~\cite[Sec.~3.6]{HASTIE-TIB-WAINWRIGHT} 
as a classifier of choice. It transforms the input features to
new representations in which the different classes are separated with margins 
that are as wide as possible. The PCA-compressed STN coordinates were fed to
the SVM classifier of a polynomial kernel of degree 1 implemented in the
\textsf{LBSVM.jl} package~\cite{LIBSVM}.
Along with the SVM, we also used the
\emph{GLMNet}~\cite[Chap.~3]{HASTIE-TIB-WAINWRIGHT} in our experiments in order
to interpret the classification results in a more intuitive manner.
The GLMNet fits a generalized linear model with Lasso regularization through
penalized maximum likelihood, and the GLMNet coefficients are denoted by
$\btheta$. The regularization path corresponding to the Lasso penalty was
computed using cyclic coordinate descent. The significance of the STN
coefficients in distinguishing the music genre can be captured by $\btheta$
when the mean loss is minimized in the \textsf{GLMNet.jl} package~\cite{GLMNet}.

\subsection{Classification results and evaluation}
We evaluated performance of various methods by comparing the predicted labels
and the ground truth of the music tracks. We computed the classification accuracy
by first computing the average accuracy under one experiment of the three-fold
cross validation, and then computing the mean of these average accuracies of
these ten repeated experiments. 

Table~\ref{tab:result} shows the superior performance of the GMW-STN with SVM
compared to the Morlet-STN with SVM. The novel incorporation of the GMWs into
the STN increased the accuracy by more than $4\%$ using the 3rd layer outputs.
Moreover, this table indicates that as the number of layers of the STN increases,
the classification accuracy also increases regardless of the wavelet filters.
The increase in accuracy is most significant from the 1st layer to the 2nd layer.

\begin{figure}
\centering
\includegraphics[scale = 0.375]{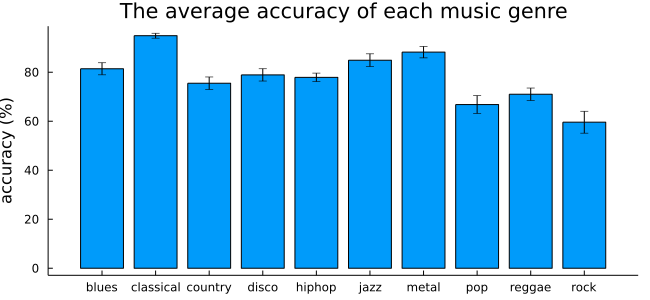}
\caption{The test accuracy of all music genres.}
\label{fig:test_acc}
\end{figure}

In comparison to the SVM classifier, the GLMNet classifier performed slightly
worse ($\sim\!\!3\%$). However, we will show in Section~\ref{sec:sig} that
the GLMNet classifier can explain the results and shed light on the music
information collected from the STN coefficients, which is impossible with the
SVM.
\begin{table}[h]
\caption{Average classification accuracy on music genres using GMW-STN and Morlet-STN}
\centering
\begin{tabular}{p{1cm}p{1.5cm}p{1.5cm}p{1.5cm}}
\hline
 &\multicolumn{2}{c}{GMW} & Morlet\\ \cline{2-4}
Layer & GLMNet & SVM & SVM\\ \hline
1  & 52.3711\% &  53.0529\% & 48.8776\% \\
  \hline
  2  & 70.2504\% &  73.7329\% & 70.0517\% \\
  \hline
  3  & 74.5500\% & {\bf 77.9088\%}  & 73.7178\%\\
\hline
\end{tabular}
 \label{tab:result}
\end{table}
Figure~\ref{fig:test_acc} displays the performance of the GMW-STN with SVM for
each individual music genre. The GMW-STN with SVM performed best in classifying
classical music $(94.9\%)$ followed by metal $(88.2\%)$, jazz $(84.9\%)$, and
blues $(81.4\%)$. However, it did not perform well in classifying
pop $(66.8\%)$ and rock $(59.6\%)$.
See Section~\ref{sec:sig} for the explanation on the difference of the
accuracies among the music genres.
Our result is comparable to the reported accuracy ($76.02\%$) without data
augmentation and ($80.93\%$) with data augmentation using
1D CNN~\cite{allamy20211d}. However, we can uniquely provide the explanation of
the results based on the additional music information retrieved in the layers,
by visualizing and interpreting the corresponding STN coefficients.
On the other hand, it is quite difficult to explain the results and interpret
the intermediate representations in deep learning.
Thus, interpretability is a main advantage of our approach over CNNs.

\section{Significance of STN Coefficients}
\label{sec:sig}
\begin{figure*}
\centering
\includegraphics[scale = 0.4]{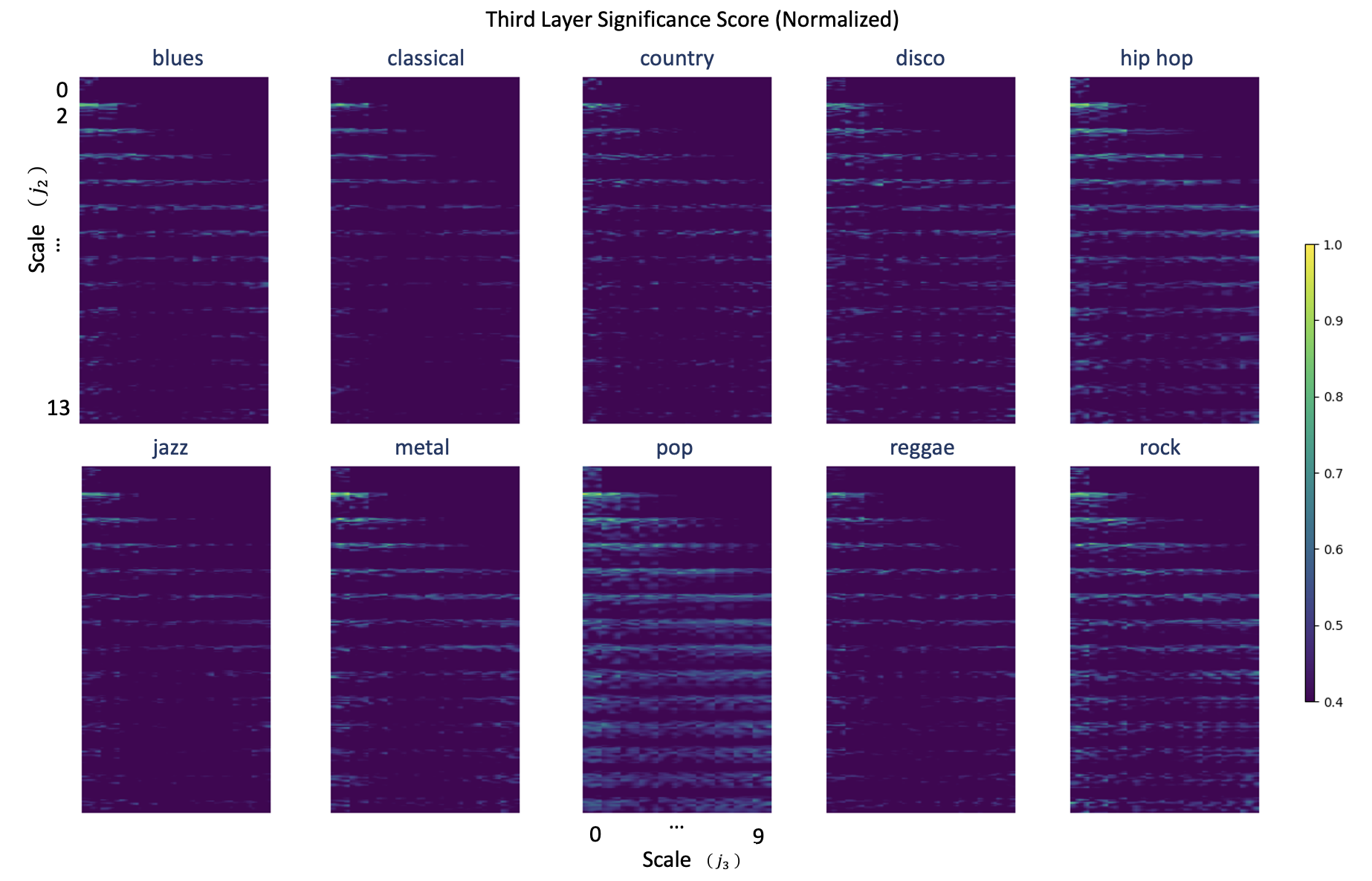}
\caption{Significance scores from the 3rd layer STN coefficients for each
  music genre}
\label{fig:score}
\end{figure*}
As we mentioned in the previous subsection, the SVM cannot
indicate which subset of the STN coefficients mainly contributed to the
correct classification results.
Unlike the SVM, the GLMNet can provide such information.
Since the GLMNet coefficient vector $\btheta$ for each
genre was computed on the top $1,000$ PCA components of the 3rd layer STN
coefficients, we first inverted the PCA to get the corresponding the
3rd layer STN coefficients. Then, we normalized these coefficients so that
the maximum value became $1$, which we call the \emph{significance scores}.
Figure~\ref{fig:score} displays these significance scores of ten music genres
with the lower bound clamped to $0.4$.
It shows that information quantified in the 3rd layer STN is critical in the
low frequency portion, especially in the $(j_3, j_2) = (0, 1)$ blocks.
In general, the concentration in the lower frequency region is positively
associated to the high classification rate.
For instance, from Figure~\ref{fig:score}, the classical music has the highest
scores in the lower frequency portion of the 3rd layer coefficients while
pop and rock have more dispersed score distributions, which can be attributed
to the variations of the music patterns in these genres, which in turn may have
contributed to the lower classification accuracies for these genres.

\section{Conclusions}
We demonstrated that the GMW-STN outperformed the conventional STN using the
Morlet wavelets. It can be explained by the importance of analyticity of
the underlying wavelet transform. In addition, the classification accuracy
became higher with more layers in the STN since we could retrieve the
more relevant music information that are stable with respect to local
deformations in the deeper layer STN coefficients.
We could illustrate the connection between the music information retrieved from
the GMW-STN and the classification results, which would be impossible using
CNNs/DNNs. In addition, it turned out that the lower frequency portion of
the music information retrieved from the 3rd layer STN coefficients mainly
attributed to the music genre classification performance.
In the near future, we plan to explore a 2D STN applied to the spectrograms of
the input music tracks.

\section*{Acknowledgments}
This research was partially supported by the US National Science Foundation grants DMS-1912747, CCF-1934568; the US Office of Naval Research grant N00014-20-1-2381.



\end{document}